\documentclass[]{spie}  

\usepackage{amsmath,amsfonts,amssymb}
\usepackage{graphicx}
\usepackage[colorlinks=true, allcolors=blue]{hyperref}

\title{Calibration of the first detector flight models for the HERMES constellation and the SpIRIT mission}

\author[a]{Riccardo~Campana}
\author[a,b]{Giulia~Baroni}
\author[c]{Giovanni~Della~Casa}
\author[d]{Giuseppe~Dilillo}
\author[a]{Ezequiel~J.~Marchesini}
\author[d]{Francesco~Ceraudo}
\author[e]{Alejandro~Guzmán}
\author[e]{Paul~Hedderman}
\author[d]{Yuri~Evangelista}
\affil[a]{INAF/OAS, Bologna, Italy}
\affil[b]{University of Bologna, Department of Physics and Astronomy, Bologna, Italy}
\affil[c]{University of Udine, Department of Mathematical, Physical and Computer Sciences, Udine, Italy}
\affil[d]{INAF/IAPS, Rome, Italy}
\affil[e]{Eberhard Karls Universität, Institute for Astronomy and Astrophysics, Tübingen, Germany}

\authorinfo{On behalf of the HERMES collaboration. \\ Further author information: send correspondence to R.C., E-mail: \url{riccardo.campana@inaf.it}}

\begin{document} 
\maketitle

\begin{abstract}
HERMES (\emph{High Energy Rapid Modular Ensemble of Satellites}) is a space-borne mission based on a constellation of six 3U CubeSats flying in a low-Earth orbit, hosting new miniaturized instruments based on a hybrid Silicon Drift Detector/GAGG:Ce scintillator photodetector system sensitive to X-rays and $\gamma$-rays. 
Moreover, the HERMES constellation will operate in conjunction with the Australian-Italian \emph{Space Industry Responsive Intelligent Thermal} (SpIRIT) 6U CubeSat, that will carry in a Sun-synchronous orbit (SSO) an actively cooled HERMES detector system payload.
In this paper we provide an overview of the ground calibrations of the first HERMES and SpIRIT flight detectors, outlining the calibration plan, detector performance and characterization.
\end{abstract}


\section{INTRODUCTION}
\label{sec:intro}  

The HERMES-Technologic and Scientific pathfinder\footnote{\url{https://www.hermes-sp.eu}} (HERMES-TP/SP)\cite{fuschino19,fiore20,evangelista20} project involves a constellation of six 3U nanosatellites hosting simple but innovative X-ray detectors for the monitoring of cosmic high energy transients, such as Gamma Ray Bursts and the electromagnetic counterparts of Gravitational Wave events. 
The main objective of HERMES-TP/SP is to prove that accurate position of high energy cosmic transients can be obtained using miniaturized hardware, with cost at least one order of magnitude smaller than that of conventional scientific space observatories and development time as short as a few years. 

The HERMES-TP project is funded by the Italian Ministry for education, university and research and the Italian Space Agency. The HERMES-SP project is funded by the European Union’s Horizon 2020 Research and Innovation Programme under Grant Agreement No. 821896. 
The constellation should be tested in orbit in $\sim$2023--24. 

HERMES-TP/SP is intrinsically a \emph{modular} experiment that can be naturally expanded to provide a global, sensitive all sky monitor for high energy transients.
As such, one HERMES detector will fly onboard the SpIRIT  (\emph{Space Industry Responsive Intelligent Thermal}) mission \cite{auchettl22}, designed by the University of Melbourne.

The foreseen detector for this experiment employs the solid-state Silicon Drift Detectors (SDD) developed by INFN and FBK in the framework of the ReDSoX Collaboration\footnote{\url{http://redsox.iasfbo.inaf.it}}. 
These devices, characterized by a very low intrinsic electronic noise, are sensitive to both X-ray and optical photons. Hence they can be exploited in the so-called ``siswich" architecture, acting both as direct X-ray detectors (the operative ``X-mode") and as photodetectors for the scintillation light produced by the absorptions of a $\gamma$-ray in a scintillator cyrstal (``S-mode").
 This allows for the realisation of a single, compact experiment with a sensitivity band from a few keV up to a few MeVs for X and $\gamma$-rays, and with a high temporal resolution ($<$µs). 

\begin{figure}[htbp]
\centering
\includegraphics[width=15cm]{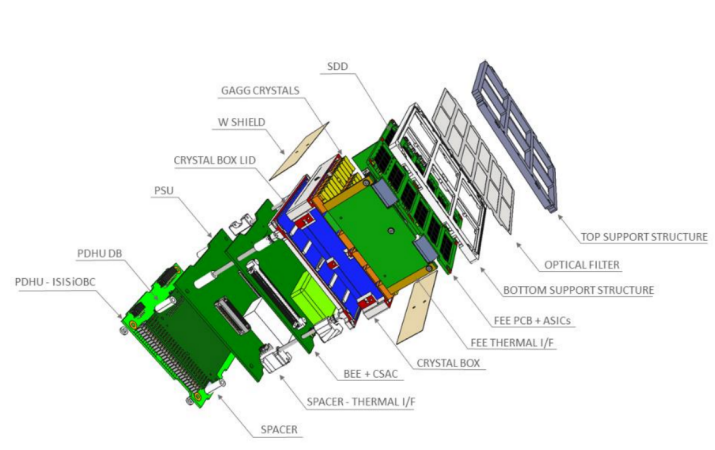}
\caption{Exploded view of the HERMES detector.}
\label{f:pl_sketch}
\end{figure}

Figure~\ref{f:pl_sketch} shows a sketch of the payload (which fits in a standard 1U CubeSat platform). 
Each of the sixty $12.10 \times 6.94 \times 15.00$ mm$^3$ scintillation crystals (made of cerium-doped gadolinium-aluminium-gallium garnet, GAGG:Ce) is coupled to two SDD channels. Ten SDD channels are integrated in a single monolithic 2$\times$5 matrix. The crystals are contained in a stainless steel box, shielded with a tungsten layer.

During the period from October 2021 to April 2022 the first flight model for the HERMES constellation (PFM, or \emph{Proto Flight Model}) and the flight model for the SpIRIT detector (FM1, or \emph{Flight Model 1}) have been integrated and calibrated.
In this paper we provide an overview of the ground calibrations of the first HERMES and SpIRIT flight detectors, outlining the calibration plan, detector performance and characterization.

\section{Detector working principle}
\label{sec:WP}  
The total number of channels in a HERMES payload is 120. Each SDD anode is connected to a front-end chipset, composed of two different ASICs.
The first one, LYRA-FE, is placed on the FEE board (Figure~\ref{f:pl_sketch}) as close as possible to the SDD anodes, and contains the preamplifier and the first stage of the shaping amplifier in a $\sim$0.5~mm$^2$ die, one for each channel. The signal is then routed to the side wing of the FEE board, to a multi-channel LYRA-BE ASIC, which handles the rest of the spectroscopic acquisition chain (second stage of the shaping amplifier, threshold discriminator, peak and hold circuit, analog output multiplexer). Each LYRA-BE handles 30 FE channels, and the whole detector is split in four totally independent quadrants.
The configuration and communication with the ASICs, and the analog-to-digital conversion of their signal, is handled by the BEE board (Figure~\ref{f:pl_sketch}), based on a Altera Cyclone V FPGA architecture.

When the signal on a given channel crosses a programmable threshold, a trigger is raised by the LYRA-BE, which then holds the signal in an above-threshold channel to its peak value. The BEE FPGA then commands the analog multiplexer, scanning first a multiplexed trigger output and identifying the triggering channel(s) address(es), and finally the analog output multiplexer is addressed and the signal converted to digital. The BEE prepares therefore an event record containing timestamp information, the address and the amplitude of each triggering channel, which is stored in a buffer and then transferred through a serial interface to the Payload Data Handling Unit (PDHU).

Usually, when an X-ray interaction in the SDD silicon bulk occurs, only one triggering channel is present (X-mode event). Otherwise, if the incoming photon is absorbed in a scintillator crystal, the two channels reading out the crystal are triggered (S-mode event).

\section{HERMES calibration overview}
The main aim of the X-mode calibration is the determination of the electronic \emph{gain} and \emph{offset} for each channel.
When an X-ray is absorbed in the SDD silicon bulk, on average one electron is produced every 3.65 eV of deposited energy (assuming as an average value the one calculated for a temperature of 300 K and at an energy of 5.9 keV \cite{mazziotta08}). Consequently, there is a linear relation between the X-ray energy and the signal amplitude in electrons.
The gain and offset are defined as the slope and intercept of the linear fit between the measurements in a recorded spectrum of known photon line energies, expressed in electrons:
\begin{equation}
    y[ADC]=x[e^- ] \times \mathrm{Gain}[ADC/e^- ]+\mathrm{Offset}[ADC]
\end{equation}
The calibration of a given recorded amplitude A in units of $[e-]$ is then performed by:
\begin{equation}
A[e^-] = \frac{A[ADC]-\mathrm{Offset}}{\mathrm{Gain}}
\end{equation}

For X-mode events, the amplitude can be therefore expressed in keV as:
\begin{equation}
A[\mathrm{keV}]=  \frac{3.65 [\mathrm{eV}/e^- ]}{1000 [\mathrm{eV/keV}]} A[e^-] = 
0.00365 [\mathrm{keV}/e^-] A[e^-] = 
0.00365 \frac{A[ADC]-\mathrm{Offset}}{\mathrm{Gain}}
\end{equation}

On the other hand, the main aim of the S-mode calibration is the verification of the \emph{effective light output} for each channel.
The effective light output $LY$ is defined as the number of electrons produced at the SDD anode for unit of energy absorbed in the scintillator crystal and is measured in $[e^-/\mathrm{keV}]$. As such, it is the product between the intrinsic light yield (i.e., the number of scintillation photons produced per unit of absorbed $\gamma$-ray energy), the optical contact efficiency (the fraction of scintillation photons able to reach the SDD surface window, which may change from one channel to another) and the SDD quantum efficiency (the fraction of scintillation photons able to produce one electron of charge in the active silicon bulk; on average, one optical photon produces an electron).
The effective light output is known to be a non-proportional quantity, i.e., its value depends also on the energy. This effect is being characterized on the same HERMES scintillator crystals with a dedicated setup (Campana et al., 2022, in preparation) and parameterized in the scientific pipeline.
The effective light output also depends on the temperature, with an expected $\sim$20\% increase by going from room temperature to $-20$ $^\circ$C.
Therefore, the signal amplitude for S-mode events is expressed in keV as:
\begin{equation}
A[\mathrm{keV}]=  \frac{A[e^-]}{LY[e^-/\mathrm{keV}]} = \frac{1}{LY[e^-/\mathrm{keV}]}   \frac{A[ADC]-\mathrm{Offset}}{\mathrm{Gain}}
\end{equation}
where $LY$ is the light output for the given channel (and the given temperature).

To reconstruct the event read out by two channels $(a, b)$ optically connected to the same crystal, its amplitude is given by:
\begin{equation}
\begin{split}
A_\mathrm{tot} [\mathrm{keV}] & = 
\frac{A_a [e^-] + A_b [e^-]}{LY_a [e^-/\mathrm{keV}] + LY_b [e^-/\mathrm{keV}]} \\
& = \frac{1}{LY_a [e^-/\mathrm{keV}] + LY_b [e^-/\mathrm{keV}]} 
\left(\frac{A_a[ADC]-\mathrm{Offset}_a}{\mathrm{Gain}_a} +
\frac{A_b[ADC]-\mathrm{Offset}_b}{\mathrm{Gain}_b} \right)
\end{split}
\end{equation}


\section{Calibration setup}
The calibration parameters to be determined are thus, for each channel, the electronic gain, the offset and the effective light output. Each of these parameters will also be temperature-dependent, therefore a calibration should be performed at different temperatures (spanning the foreseen operating temperature, which is $\sim5$ $^\circ$C for HERMES and $\sim -15$ $^\circ$C for SpIRIT).
The calibration is performed by placing the flight unit in a suitable temperature-controlled climatic chamber, and illuminating the detector with standard laboratory calibration sources (e.g., $^{55}$Fe, $^{109}$Cd, $^{241}$Am, $^{137}$Cs).

\begin{figure}[htbp]
\centering
\includegraphics[width=0.9\textwidth]{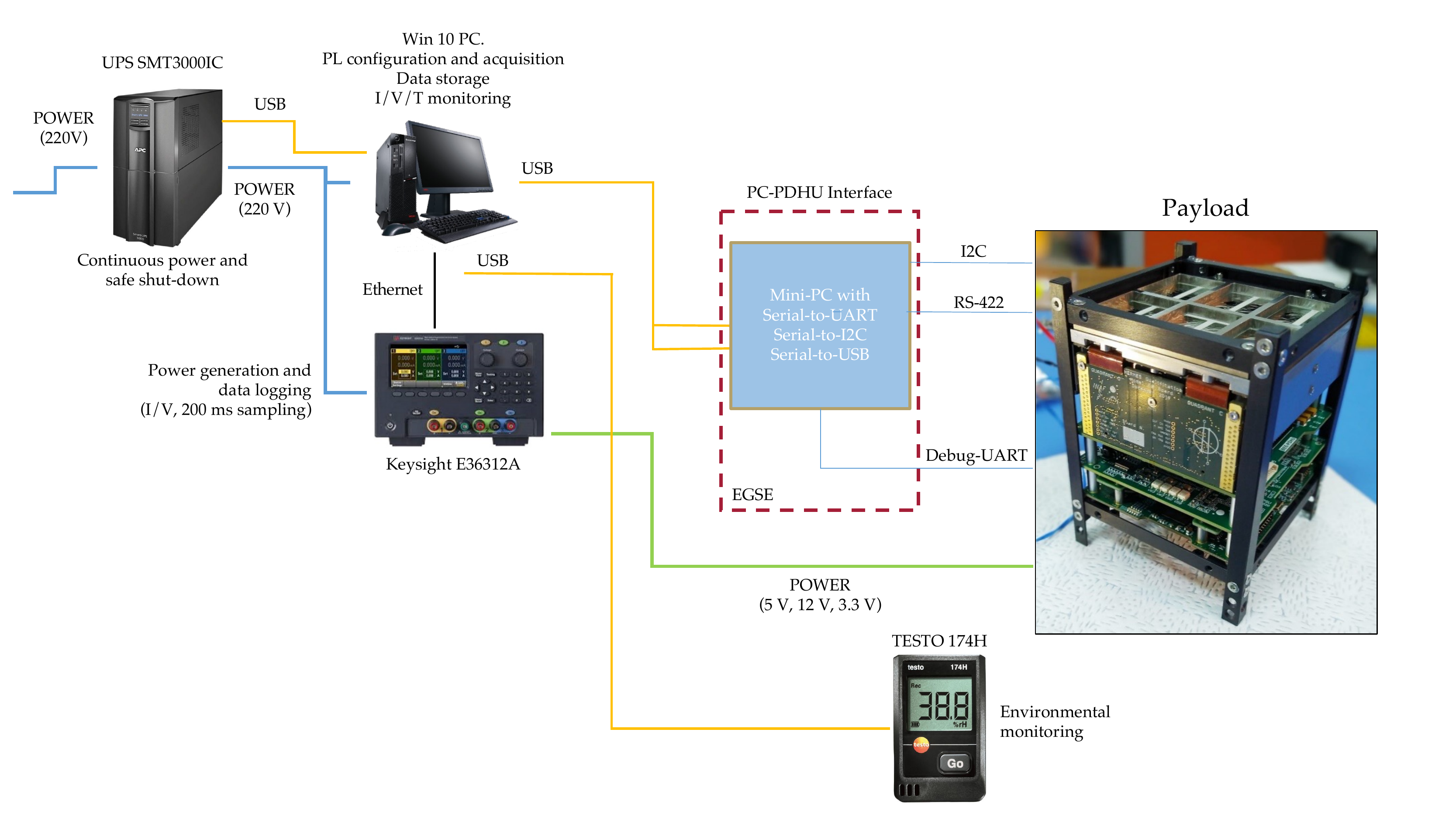}
\caption{Sketch of the HERMES EGSE used for the calibrations.}
\label{f:egse}
\end{figure}

The HERMES calibrations are performed in a class 10000 clean room at the INAF/IAPS laboratories in Rome, Italy.
The electronic ground segment equipment (EGSE) used  is shown in Figure~\ref{f:egse}. The flight payload unit is placed in the climatic chamber, and the PDHU is connected to a standard laboratory PC using a custom-made interface. A desktop power supply provides the three power lines required by the payload (3.3~V, 5.0~V and 12~V).

The  climatic chamber has an available internal volume of around 1000~cm$^3$.
The radioactive sources are hosted on a custom-made, 3-D printed source holder. Suitable recesses allow to host the available radioactive sources, using also specific adapters. A 3-D printed cover allows to fix the sources in place. The source holder is mounted on a movable arm placed inside the climatic chamber, with the sources usually placed at 18--20~cm from the detector top side, allowing for an almost uniform coverage of the detector plane (maximum off-axis angle is $\sim$15$^\circ$--20$^\circ$).

The calibration plan executed on the PFM and FM1 flight units can be summarized as follows:
\begin{enumerate}
    \item Radioactive sources ($^{55}$Fe, $^{109}$Cd, $^{241}$Am, $^{137}$Cs) are placed on the source holder. Usually, a measurement is performed using several sources at once.
    \item The climatic chamber is set at a given temperature. The temperature steps used are +20~$^\circ$C, +10~$^\circ$C, 0~$^\circ$C, $-$10~$^\circ$C, $-$20~$^\circ$C.
    \item The payload is switched on and put in the \emph{Idle} operative mode (all the quadrants configured and detector high voltage ramped up).
    \item Once the payload is in a stable thermal environment (as monitored by the external and internal temperature sensors) a data acquisition is 
    performed, accumulating at least $10^5$ counts for each radioactive source and each channel.
    \item Data is downloaded and analyzed.
    \item The climatic chamber is set at an another temperature, and the sequence starts again.
\end{enumerate}

\begin{figure}[htbp]
\centering
\includegraphics[width=7cm]{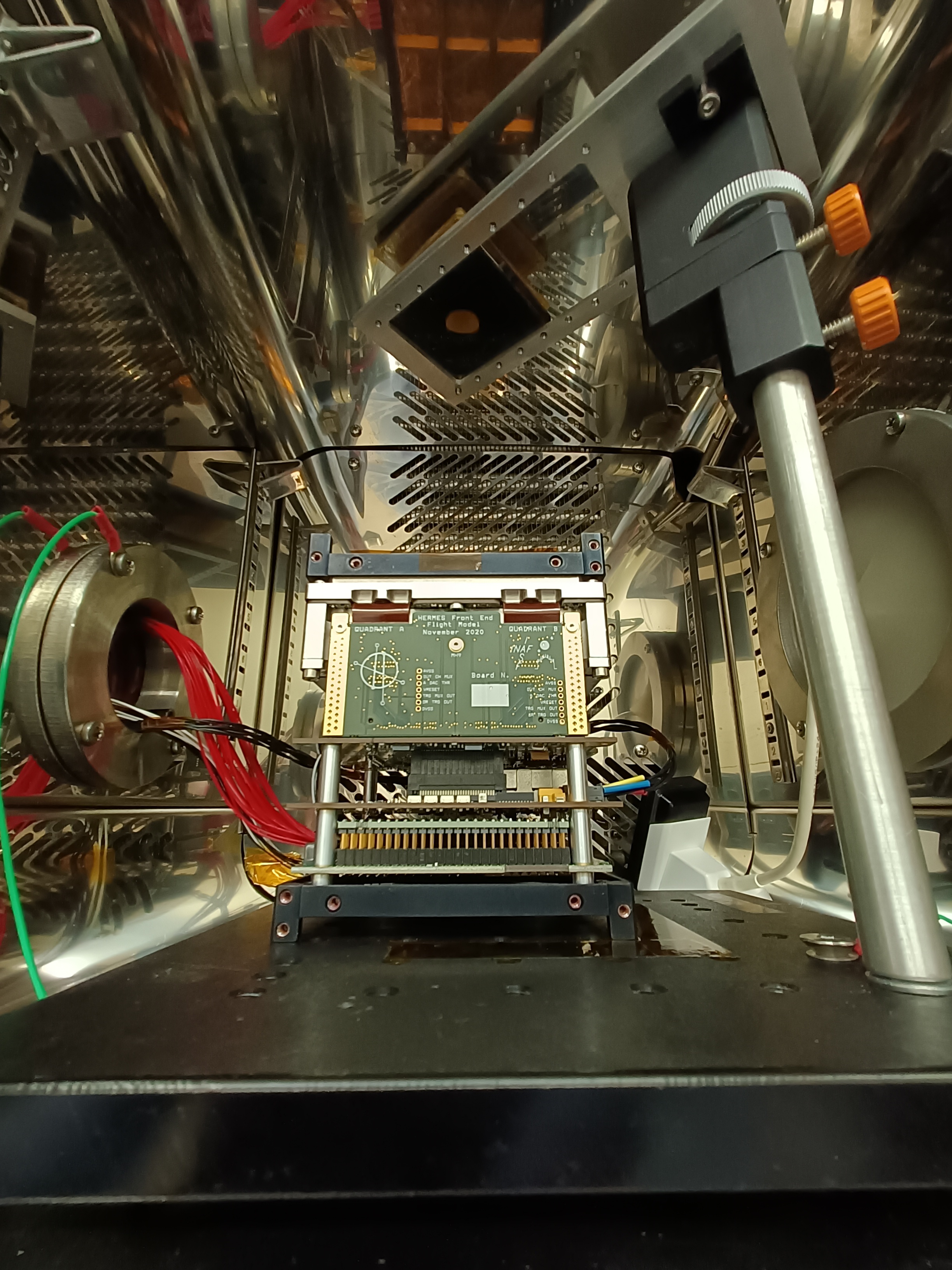}
\caption{HERMES FM1 unit placed inside the climatic chamber. The source holder is visible on the upper side.}
\label{f:hermescal}
\end{figure}

\section{Data reduction pipeline}
The raw, binary data acquisition files are converted in a standard FITS ``Level 0'' format. An additional conversion in a FITS ``Level 0.5'' format useful for further processing is also performed.
The data calibration and analysis is performed using \texttt{mescal}, a custom-made highly automatized pipeline of Python scripts.

The \texttt{mescal} pipeline (Figure~\ref{f:mescal}) requires a Level 0.5 input file, and to explicitly define the list of the radioactive sources which were used for the calibration data acquisition. The given energy units for the calibration source list will be the units of the output calibrated spectra. Then, the general algorithm proceeds as follows:

\begin{figure}[htbp]
\centering
\includegraphics[height=6.25cm]{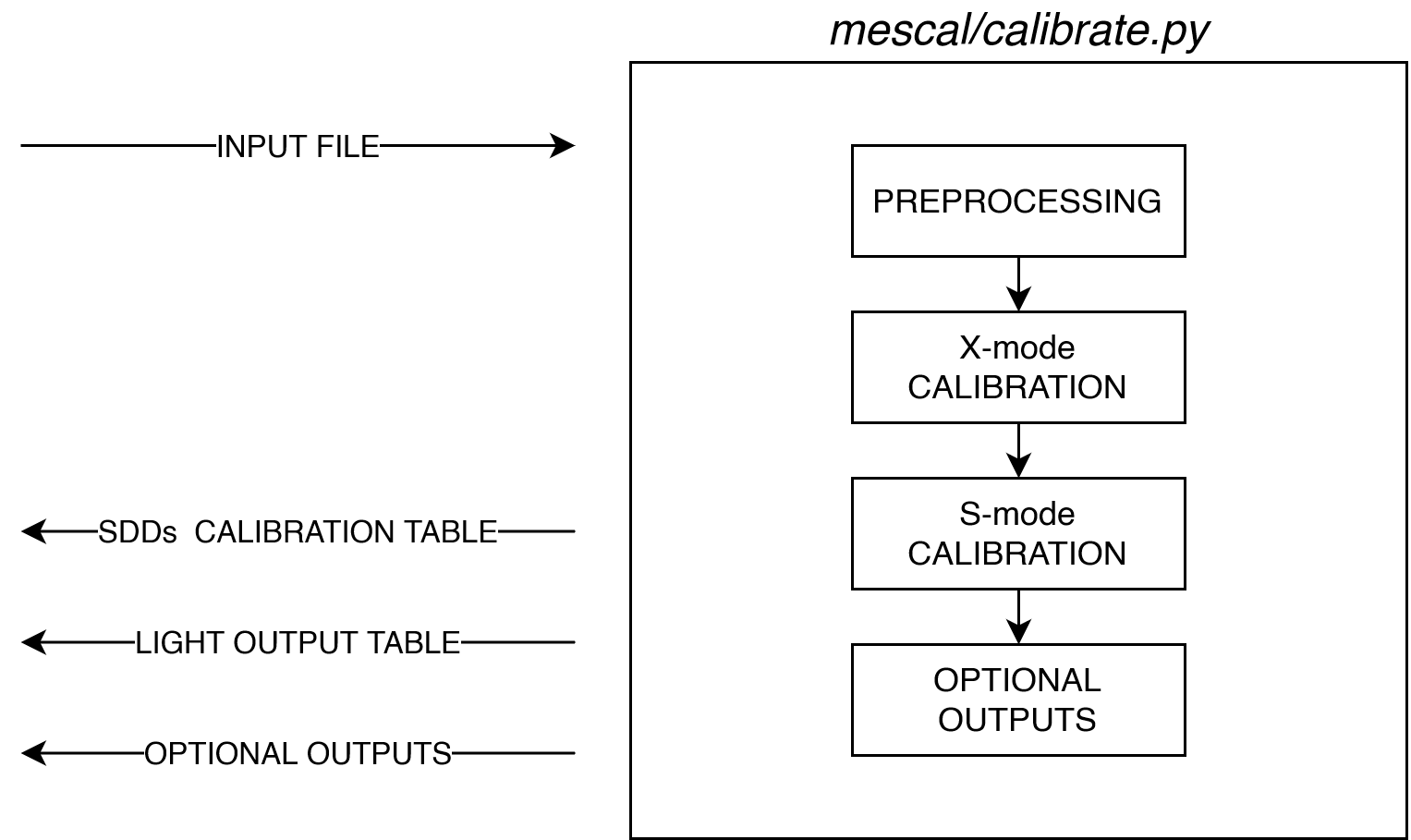}\hfill
\includegraphics[height=6.25cm]{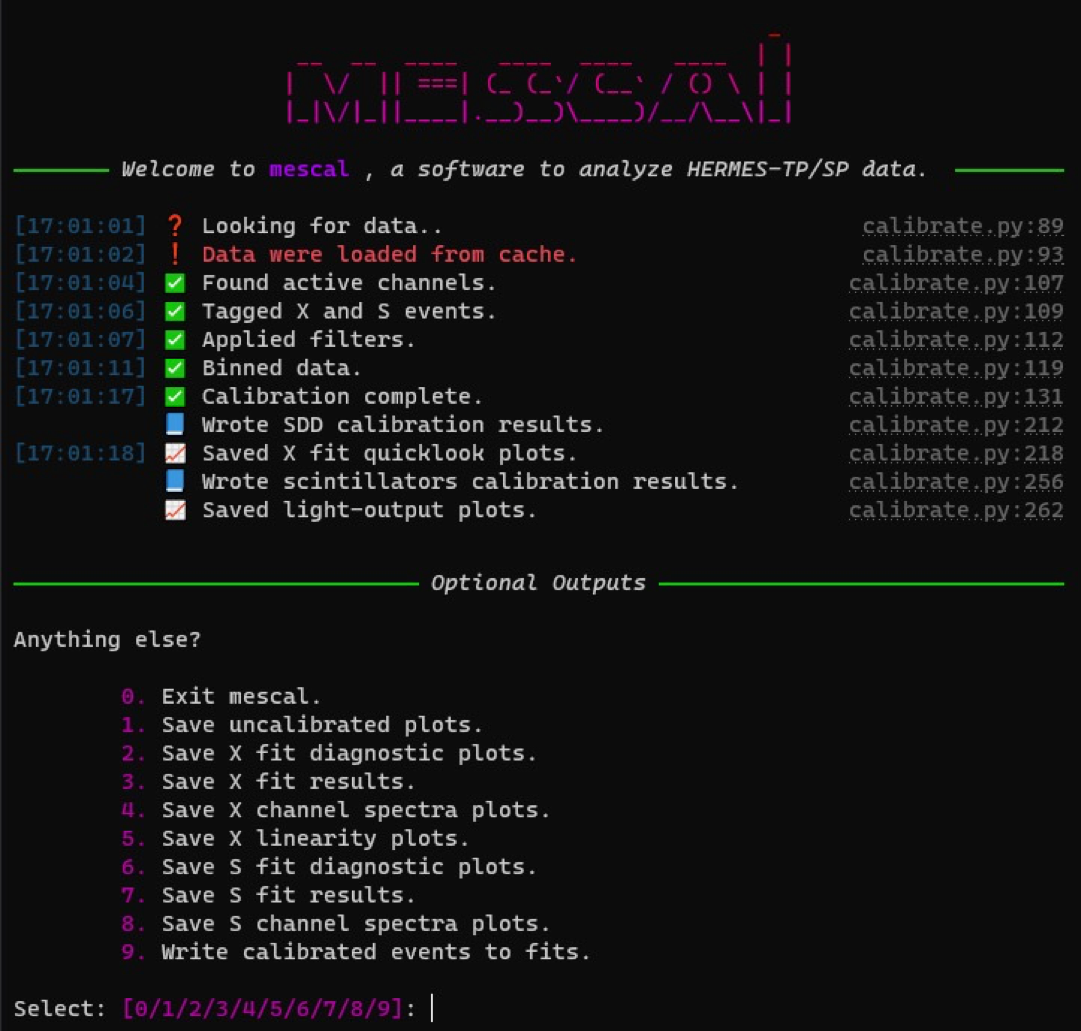}
\caption{Left: block diagram of the \texttt{mescal} logic structure. Right: output option menu given by \texttt{mescal}.}
\label{f:mescal}
\end{figure}

\begin{enumerate}
    \item \emph{Data readout} \\
    The input data is readout and organized in a Python \texttt{pandas} data frame, i.e.,  a table with rows representing a photon event and columns storing the relevant information.   
    
    \item \emph{Event type discrimination and tagging, multiplicity filtering and histogram building} \\
    Events are tagged as ``X-mode'' or ``S-mode'', to discriminate between events generated directly in the SDDs and events generated in the scintillators. The logic behind the event discrimination process consists of looking at the list of the channels readout in the same trigger cycle, and tag events seen simultaneously by both channels reading the same crystal as `S', otherwise as `X'. Tags are defined and stored for each individual event. It is worth noting that this tag does not consider event multiplicity. Two spectra are then obtained, by building the histogram of `X'- and `S'-tagged events, respectively. These spectra only include events with multiplicity of exactly 1, for the former, and with multiplicity of exactly 2 for the latter.
   
    \item \emph{X-mode calibration}\\
    If at least one of the user-input radioactive sources is listed as a X-ray emitting calibration source, then the algorithm will attempt to perform energy calibration on the X-mode spectra. To achieve this, the relative differences between all known X-ray lines of all the radioactive sources are measured, in the energy space. 
    Then, a peak-finding filter is run on the data. The peaks are chosen by comparing each local maxima to their immediate surroundings, defining a ranking of prominence for each line. These positions are used as input for a Gaussian-profile fitting algorithm (using the \texttt{lmfit} package), which returns the best-fit peak position and full-width half maximum (FWHM) for each line.
    
    The differences between these peak positions in the ADC space and in the energy space for each line are then compared, for different sets of $N$ data lines, where $N$ is the number of all the X-mode calibration lines that are being used. The set of $N$ data lines with the smallest distance dispersion with respect to the calibration lines will be chosen, and identified with the respective calibration lines.

    The set of pairs composed of the ADC value for the centroid of a data line and the energy value of its corresponding calibration line are given as input to a least-squares linear fit algorithm. The fit will return gain, offset and their errors for each channel. 
    The console will list for which channels, if any, this calibration step failed.
    
    Each event will then be separately calibrated for each channel in energy units, and a final calibrated X-mode event list will be created.
    
    \item \emph{S-mode calibration}\\
    If at least one of the user-input radioactive sources is listed as a $\gamma$-ray emitting calibration source, the algorithm will attempt to perform energy calibration on the S-mode spectra. Since the scintillator effective light output is referenced to the 661.67~keV $^{137}$Cs line, the algorithm automatically uses the peak finding procedure to find this line, assuming it is the only line present. Then, a Gaussian-profile fitting algorithm is run on the position given by the peak finder, which returns the position of the center of the line and its FWHM.
    
    Then, for each channel, the algorithm takes the gain and offset parameters found during the X-mode calibration and converts the S-mode spectrum in units of electrons. This spectrum is then used to derive the light output for each single channel by fitting the  661.67~keV line.
    
    Finally, for each pair of channels reading the same crystal, the events are summed event-by-event in electrons and normalized to the sum of the light output of both coupled channels. This returns a crystal-summed event list in energy units. 
    The console will list the channels, if any, for which this calibration step failed.

    \item \emph{Visualization plots and output writing}\\
    The program will ask the user to define what results to write as output tables, and which plots to be saved as output (Figure~\ref{f:mescal}). 
\end{enumerate}

\section{Overview of the results}

\begin{figure}[htbp]
\centering
\includegraphics[height=5cm]{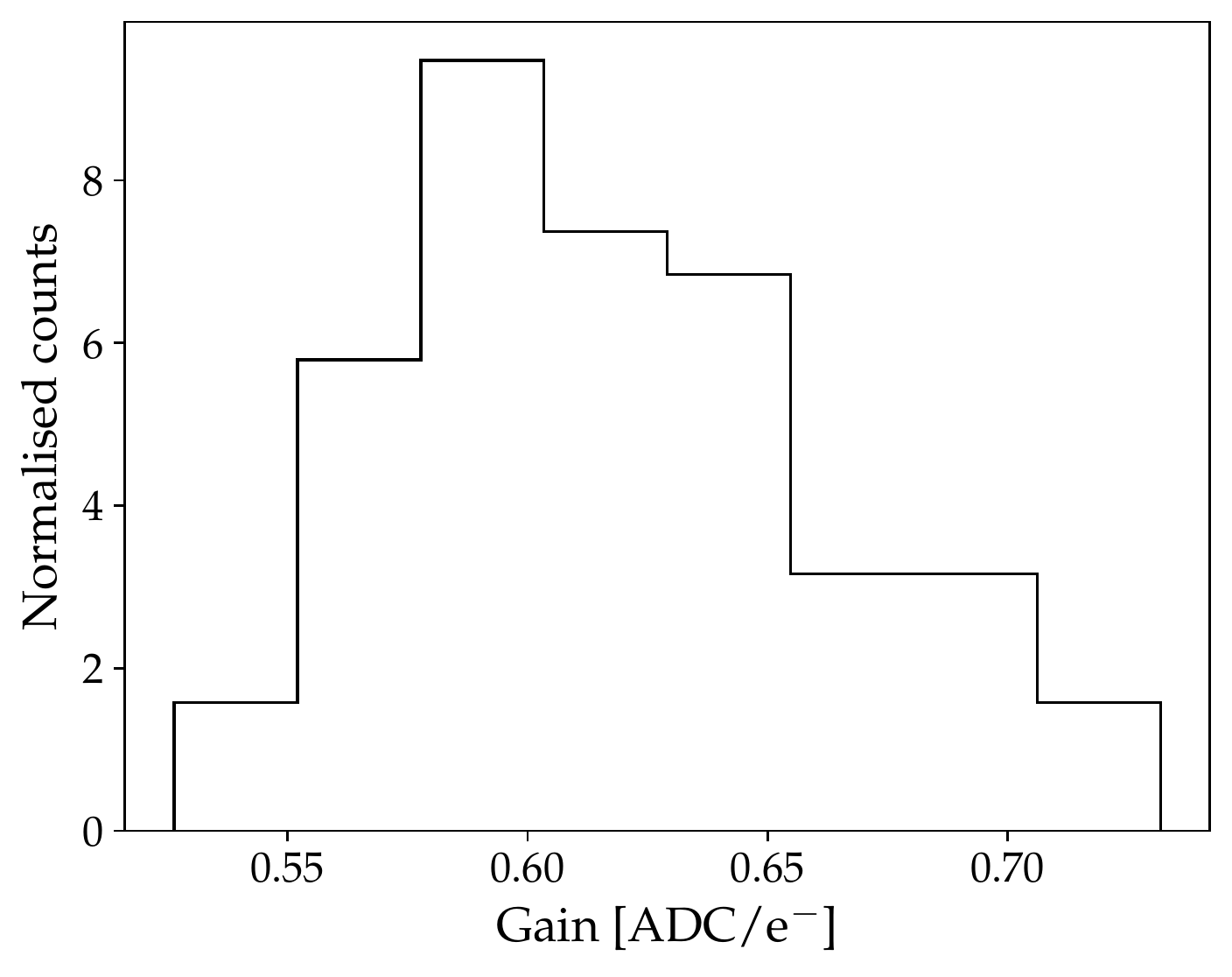}
\includegraphics[height=5cm]{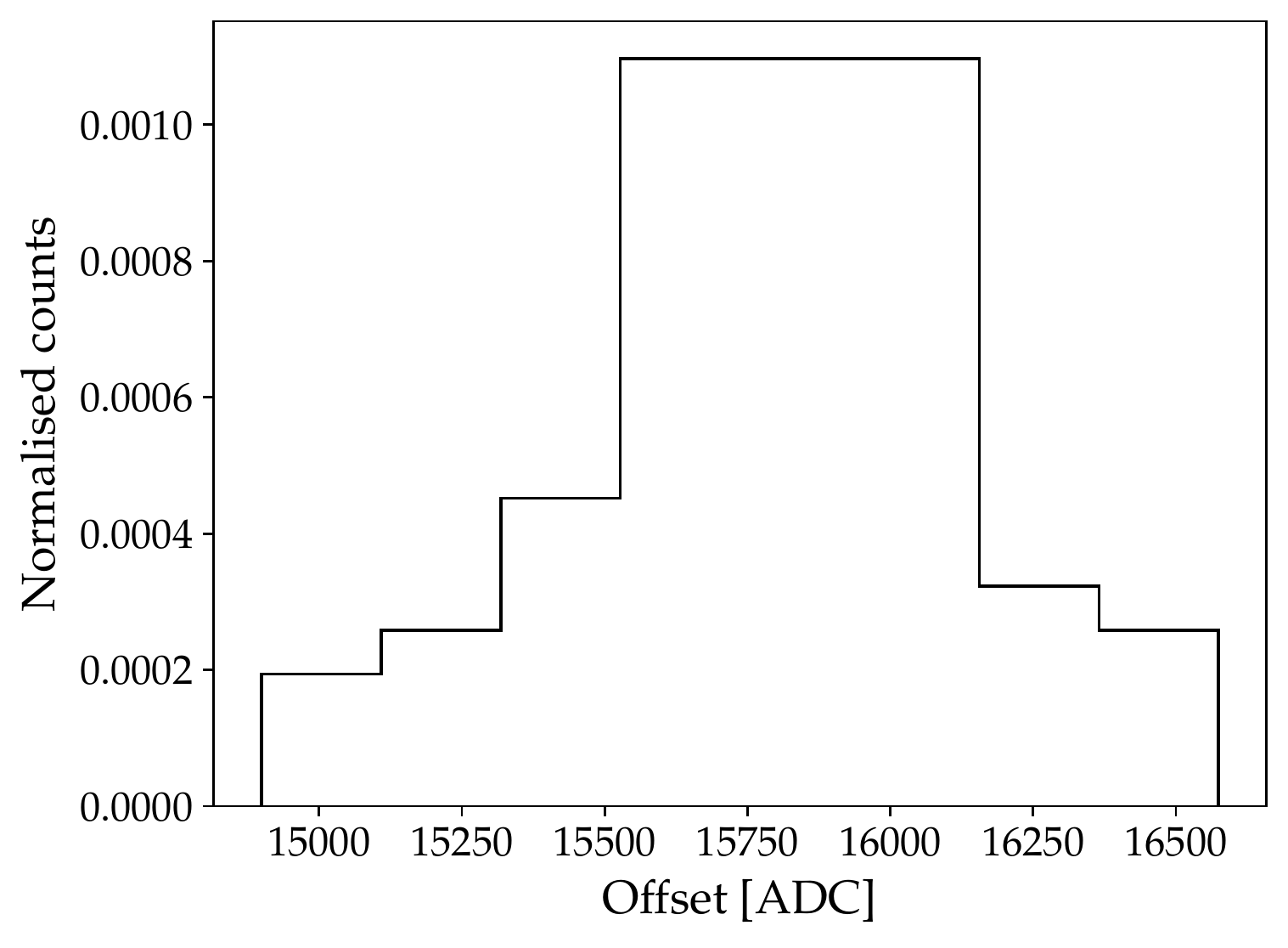}
\caption{Distribution of the measured gain and offset values on the HERMES FM1 payload, at a temperature of $-$10~$^\circ$C.}
\label{f:gain_offset}
\end{figure}

Figure~\ref{f:gain_offset} shows the measured distribution of gain and offset on the FM1 channels. The spread around the average value is of a few percent, and consistent with the expectations.

\begin{figure}[htbp]
\centering
\includegraphics[height=7cm]{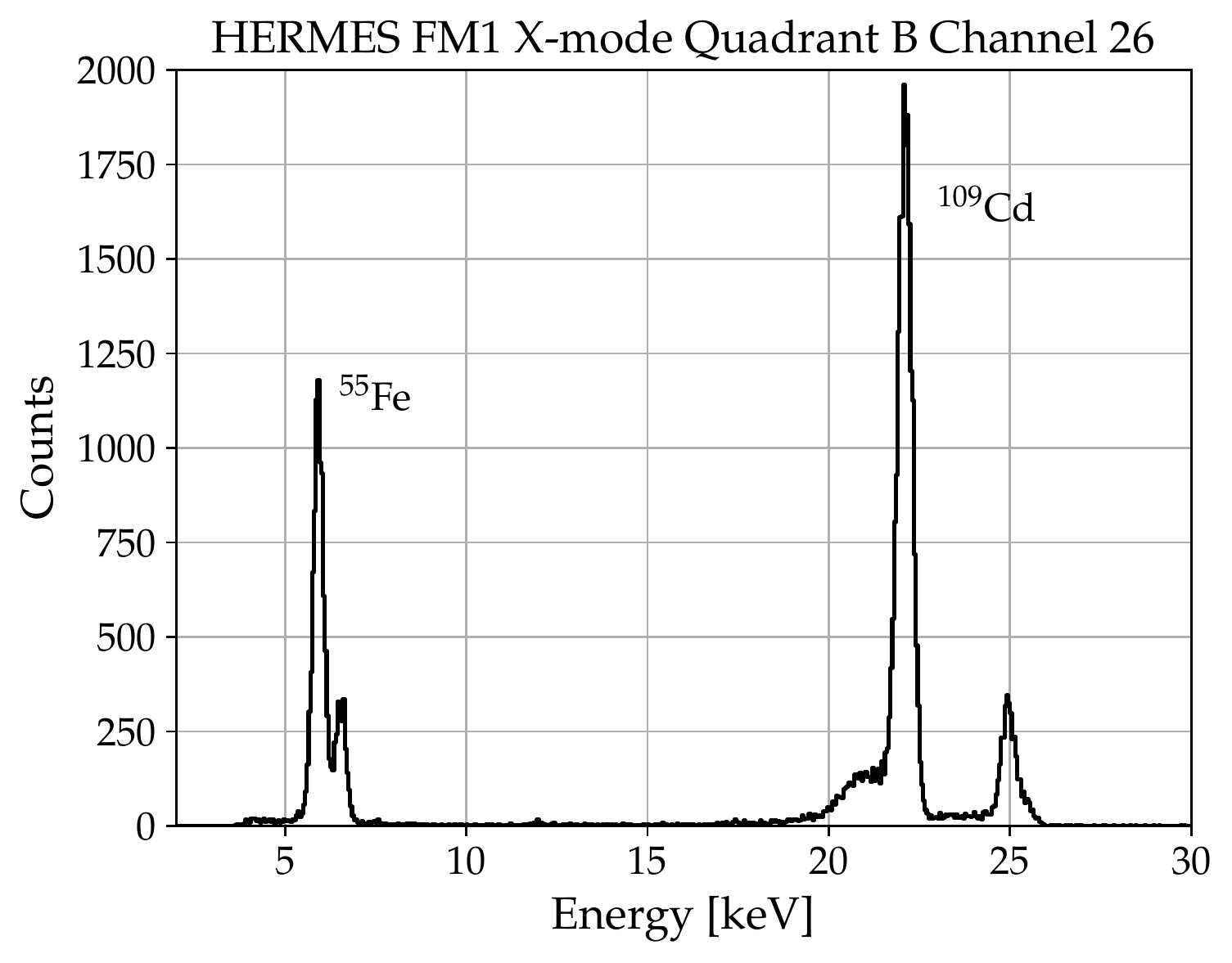}
\caption{X-mode calibrated spectrum of $^{55}$Fe and $^{109}$Cd radioactive sources for a representative channel, at a temperature of $-$20~$^\circ$C.}
\label{f:xmodespectrum}
\end{figure}

Figure~\ref{f:xmodespectrum} shows the calibrated X-mode spectrum for a representative channel, showing the X-ray lines of the $^{55}$Fe and $^{109}$Cd sources between $\sim$6 and $\sim$25 keV. The lower threshold is below 2 keV, while the resolution at 6 keV is $\sim$300 eV FWHM.

\begin{figure}[htbp]
\centering
\includegraphics[height=7cm]{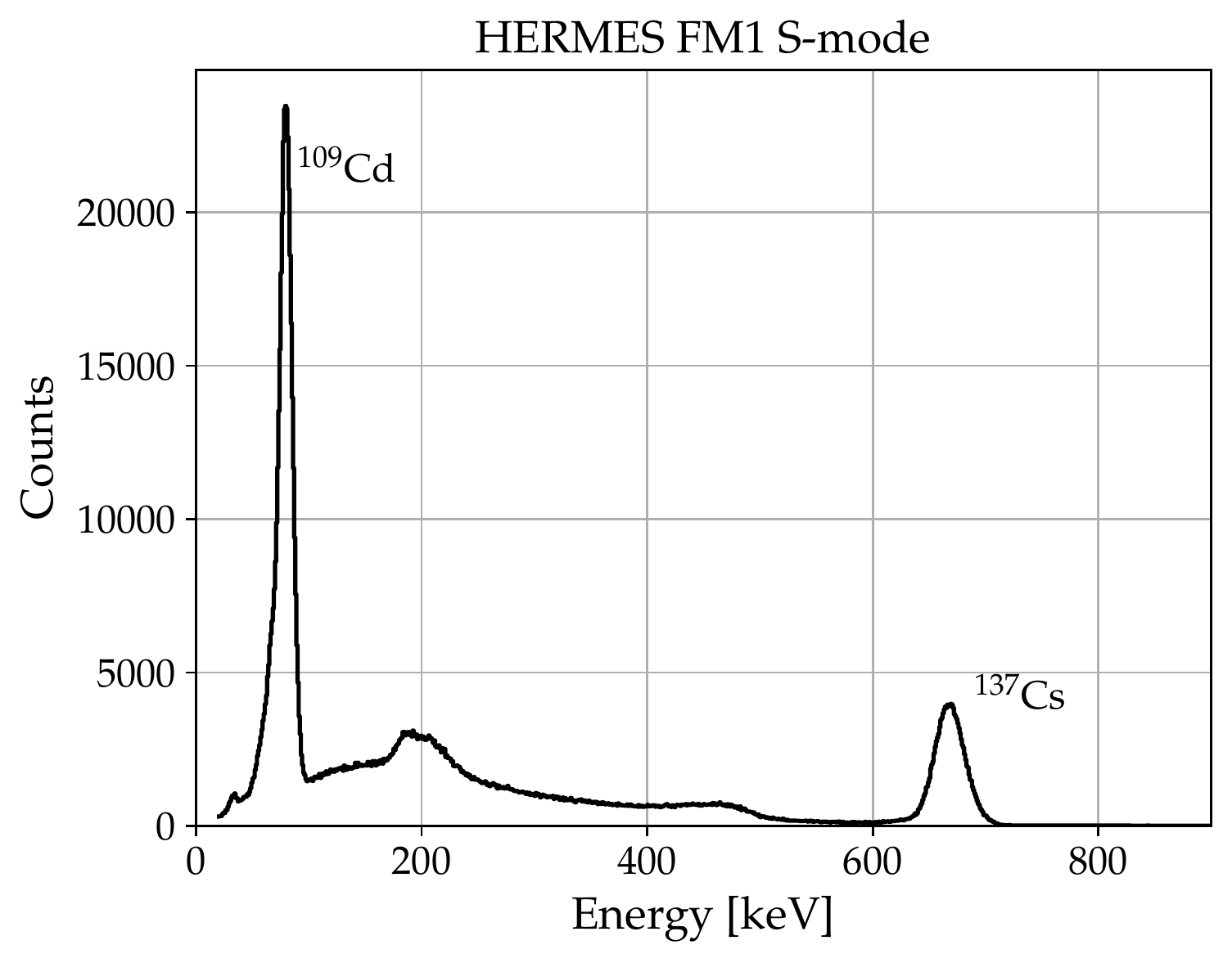}
\caption{Summed S-mode calibrated spectrum of the whole FM1 HERMES detector, illuminated with $^{137}$Cs and $^{109}$Cd radioactive sources, at a temperature of $-$20~$^\circ$C. The Cd 88 keV and Cs 662 keV lines (with Compton continuum) are well apparent.}
\label{f:smodespectrum}
\end{figure}

Figure~\ref{f:smodespectrum} shows the S-mode spectrum for the whole detector, calibrated and summed over all the channels and scintillator crystals, showing the $\gamma$-ray lines of the $^{137}$Cs and $^{109}$Cd sources at $\sim$662 and $\sim$88 keV. 

\section{Conclusions}
The first two HERMES flight models, PFM and FM1 (SpIRIT), have been successfully integrated and calibrated. For each of the 120 detector channels, characteristics such as the gain, offset and effective light output have been determined, also as a function of the operating temperature.

The FM1 detector is presently being integrated with the SpIRIT spacecraft at the University of Melbourne (Australia). The PFM will be integrated with the HERMES platform in September 2022 at Polytechnic of Milan. At each stage of the full satellite integration suitable measurements will be performed to check and verify the validity of the calibrations.

The next five HERMES flight models (FM2 to FM6) will be integrated and calibrated in the second half of 2022.

\acknowledgments 
 This work has been carried out in the framework of the HERMES-TP and HERMES-SP collaborations. We acknowledge support from the European Union Horizon 2020 Research and Innovation Framework Programme under grant agreement HERMES-Scientific Pathfinder n. 821896 and from ASI-INAF Accordo Attuativo HERMES Technologic Pathfinder n. 2018-10-H.1-2020.

\bibliography{report} 
\bibliographystyle{spiebib} 

\end{document}